\documentclass[12pt]{article}

\usepackage{jheppub}
\usepackage{amsbsy,amsfonts,amsmath,amssymb}
\usepackage{cleveref} 
\usepackage[g]{esvect} 
\usepackage{makecell}

\crefname{equation}{}{}
\crefname{chapter}{Chapter}{Chapters}
\crefname{section}{Section}{Sections}
\crefname{subsection}{Subsection}{Subsections}
\crefname{subsubsection}{Subsubsection}{Subsubsections}
\crefname{figure}{Figure}{Figures}
\crefname{table}{Table}{Tables}
\crefname{appendix}{Appendix}{Appendices}

\allowdisplaybreaks

\begin{document}

\title{Forced non-conformal relativistic fluid from the Chamblin-Reall gravity}

\author[]{Chao Wu, Hao Hu, Ruohan Wang and Tingqing Zhou}

\affiliation[]{School of Physics and Optoelectronic Engineering, Anhui University, Hefei 230601, China}

\emailAdd{chaowu@ahu.edu.cn, hao.hu@student.kit.edu, \\ wangrh67@mail2.sysu.edu.cn, zhoutq2730@stu.ahu.edu.cn}

\abstract{The Chamblin-Reall gravity is a remarkable non-conformal platform for the fluid/gravity correspondence to achieve its maximum efficiency. When a probe scalar field that does not change the background metric is manually introduced into the action of the gravity side, an external scalar field will appear at the boundary, and the gradients of the external scalar field will act as a driving force exerting on the dual relativistic fluid. Thus the dynamics of the fluid will be affected in the way that the stress tensor is no longer conserved. We will use the fluid/gravity correspondence to derive the transport coefficients related to the external scalar field and the explicit expression of the driving force.}


\arxivnumber{2412.01146}

\maketitle

\section{Introduction}

Studies on the non-conformal relativistic fluid have been going on steadily these days. With the conclusions of \cite{Wu2111}, we know that the rigorous and analytic results of the transport coefficients for non-conformal fluids can be extracted from the Chamblin-Reall gravities with one scalar field.
These Chamblin-Reall backgrounds are the $(p-q+2)$-dimensional reduced AdS black hole, the $(p+2)$-dimensional reduced Dp-brane, the $(p-q+2)$-dimensional reduced compactified Dp-brane, and the $(p+2)$-dimensional reduced smeared Dp-brane, with $q$ the number of compactified directions.
Their higher dimensional origins are separately the $(p+2)$-dimensional compactified AdS black hole, the Dp-brane, the compactified Dp-brane, and the smeared Dp-brane.

The compactified Dp-brane and the smeared Dp-brane give the same results of the transport coefficients since they are connected by T-dual \cite{Wu2307}. Thus the independent Chamblin-Reall models with one scalar field are the reduced AdS black hole, the reduced Dp-brane and the reduced compactified Dp-brane. Based on previous works \cite{Wu1508,Wu1604,Wu1807,Wu2012,Wu2111,Wu2307}, we already know well about the second-order dynamical transport coefficients for the above-mentioned gravitational backgrounds.

The calculation approach that is used in \cite{Wu1508,Wu1604,Wu1807,Wu2012,Wu2111,Wu2307} is the fluid/gravity correspondence, whose original framework is developed for a 5-dimensional AdS black hole \cite{Bhattacharyya0712,Bhattacharyya0803}. But \cite{Bhattacharyya0712} preset the boundary metric to be flat, so it misses the second-order transport coefficient $\kappa$ which is in connection with the curvature of the spacetime the fluid lives in. Soon after that, $\kappa$ of the 5-dimensional AdS black hole is uncovered in \cite{Bhattacharyya0806} by setting the metric of the boundary fluid to be curved.

Another interesting achievement of \cite{Bhattacharyya0806} is that the authors manually add a scalar field in the bulk. The manually added scalar does not change the gravitational background, for which we can see it as a probe in the bulk. The scalar probe varies slowly in the boundary directions so that it can be expanded via derivatives of the boundary coordinates just like the bulk metric. The boundary limit of the bulk scalar probe is an external scalar field depending mildly on the boundary coordinates. The gradients of the external scalar act as a driving force exposes to the relativistic fluid. \cite{Bhattacharyya0806} studies the 4-dimensional forced conformal fluid that is dual to the AdS$_5$ black hole with a manually added scalar probe. This study is generalized into various dimensions by using the AdS black hole of general dimensions with a scalar probe \cite{Ashok1309}.

To see the statements of the previous paragraph more rigorously, we consider a $d$-dimensional relativistic fluid in the curved background, whose action can be written as $S = S_g + S_M$, here $S_g$ and $S_M$ are the gravity and the matter action, respectively. To be specific, they are
\begin{align}
  S_g = \frac{1}{2\kappa_d^2} \int d^d x \sqrt{-g} R[g], \qquad S_M = \int d^dx \sqrt{-g} e^{-\phi} \mathcal L[g_{\mu\nu}, T, u^\mu, \phi].
\end{align}
In $S_M$, $T$ and $u^\mu$ are the temperature and the speed vector of the $d$-dimensional relativistic fluid, which are also called the fluid fields. $\phi$ and $g_{\mu\nu}$ can be seen separately as the external scalar field and external gravity field coupled to the fluid. In the fluid/gravity correspondence, $\phi$ is the boundary value of the bulk scalar probe $\Phi$. Under a translation of the coordinates $x^\mu\to x^\mu + \delta x^\mu = x^\mu + \varepsilon^\mu$, the external scalar field varies as $e^{-\phi(x)} \to e^{-\phi(x+\varepsilon)} = e^{-\phi} (1 - \nabla_\mu \phi \varepsilon^\mu)$, thus the variation of $S_M$ is
\begin{align}
  \delta S_M = \int d^dx \sqrt{-g} [\nabla_\nu T^\nu_{\ \mu} - e^{-\phi} \mathcal L \nabla_\mu \phi] \varepsilon^\mu.
\end{align}
So $\delta S_M/ \varepsilon^\mu = \delta S_M/ \delta x^\mu = 0$ gives the conserved current equation for $T^{\mu\nu}$ as
\begin{align}\label{eq: the forced NS eq}
  \nabla_\mu T^{\mu\nu} = f^\nu = e^{-\phi} \mathcal L \nabla^\nu \phi.
\end{align}
This is also the forced Navier-Stokes equation. From the above one can infer that an external scalar field coupled to the fluid fields prevents the stress-energy tensor from being conserved. The reason behind the non-conservation of the stress tensor is that the translational invariance of the boundary is broken by the appearance of the external scalar $\phi(x^\mu)$.

The fluid/gravity correspondence can help us to nail down the explicit form of $e^{-\phi} \mathcal L$ via the derivative terms of $\phi$, and one can easily get the force $f^\mu$ by multiplying $\nabla^\mu \phi$ to $e^{-\phi}\mathcal L$ once it is known. So in the next sections, we will only mention how to get the Lagrangian density $e^{-\phi}\mathcal L$ but not the force $f^\mu$. This article adopts the same method as \cite{Bhattacharyya0806,Ashok1309}, but uses it in a much more general stage: the non-conformal Chamblin-Reall gravity. In the next three sections, we will separately calculate the second-order transport coefficients related to the external scalar field and find the explicit form of the Lagrangian density for the reduced AdS black hole, the reduced Dp-brane, the reduced compactified Dp-brane, and the reduced smeared Dp-brane.

\section{The reduced compactified AdS black hole}

\subsection{The background and scalar probe}

The total action for the reduced AdS black hole in the Chamblin-Reall form \cite{Wu2111} with the manually added scalar probe can be written as
\begin{align}\label{eq: action in CR form rAdS}
  S =&\, \frac1{2\kappa_{p-q+2}^2} \int d^{p-q+2}x \sqrt{-g} \left[ R - \frac12 (\partial \varphi)^2 - V(\varphi) - \frac12 (\partial \Phi)^2 \right] \cr
  & - \frac1{\kappa_{p-q+2}^2} \int d^{p-q+1}x \sqrt{-h} \left[ K + \frac{p}{L} \, e^{- \frac{\gamma}{2} \varphi}  - \frac{L}{4(p-1)} e^{\frac\gamma2 \varphi} (\bar \partial \Phi)^2 \right],
\end{align}
where the potential of the scalar field $\varphi$ and the parameter $\gamma$ are
\begin{align}
  V(\varphi) = - \frac{p(p+1)}{L^2} e^{- \gamma \varphi}, \qquad \gamma^2 = \frac{2q}{p(p-q)}.
\end{align}
The scalar field $\varphi$ comes from the dimensional reduction so we can call it the dilaton. $\Phi(x^M)$ is a scalar field introduced by hand. It does not change the bulk metric so we call it the probe scalar field, or just scalar probe.

The equations of motion (EOMs) derived from \cref{eq: action in CR form rAdS} are
\begin{align}
  & E_{MN} - T_{MN} = 0, \label{eq: Einstein equation} \\
  & \nabla^2 \varphi - V'(\varphi) = 0, \label{eq: EOM varphi}  \\
  & \nabla^2 \Phi = 0, \label{eq: EOM Phi}
\end{align}
where $E_{MN} = R_{MN} - \frac12 g_{MN}R$ and
\begin{align}\label{eq: bulk energy-momentum tensor}
  T_{MN} = \frac12 \Big(\partial_M \varphi \partial_N \varphi - \frac12 g_{MN} (\partial \varphi)^2\Big) - \frac12 g_{MN} V(\varphi) + \frac12 \Big(\partial_M \Phi \partial_N \Phi - \frac12 g_{MN} (\partial \Phi)^2 \Big)
\end{align}
are separately the Einstein tensor and the energy-momentum tensor in the bulk, and $V'(\varphi) $ for the present case is just
\begin{align}
  V'(\varphi) = \frac{dV(\varphi)}{d \varphi} = \frac{p(p+1)}{L^2} \gamma e^{- \gamma \varphi}.
\end{align}
The background that solves the EOMs of the reduced compactified AdS black hole, i.e. \cref{eq: Einstein equation,eq: EOM varphi,eq: EOM Phi}, is
\begin{align}\label{eq: background solution rAdS}
  ds^2 =& \left( \frac r{L} \right)^{\frac{2p}{p-q}} \left( -f(r) dt^2 + d \vv x^2 + \frac{L^4 dr^2}{r^4 f(r)} \right), \qquad e^\varphi = \left( \frac{r}{L} \right)^{p \gamma}, \qquad \Phi = \phi_0.
\end{align}
Here $f(r) = 1 - r_H^{p+1}/r^{p+1}$ is the emblackening factor and $\phi_0$ is a constant. We set the coordinates as $x^M = (x^\mu,r) = (t, x^i, r)$, where $x^i$ has $p-q$ components in the cases of reduced AdS black hole and the reduced compactified Dp-brane, and it has $p$ components for the cases of reduced Dp-brane and the reduced smeared Dp-brane.

After the boost transformation on the metric and letting all the fluid parameters $r_H$ and $u^\mu$ be $x^\mu$-dependent, we need to also include all the metric perturbations to make the metric satisfy the EOMs, what we have is the global on-shell metric:
\begin{align}\label{eq: global on-shell metric rAdS}
    ds^2 =&\; r^\frac{2p}{p-q} \bigg[ - \Big( f(r_H(x),r) + k(r_H(x), u^\alpha(x), r) \Big) u_\mu(x) u_\nu(x) dx^\mu dx^\nu \cr
    & - 2 P_\mu^\rho(u^\alpha(x)) w_\rho(u^\alpha(x), r) u_\nu(x) dx^\mu dx^\nu  \cr
    & + \Big( P_{\mu\nu}(u^\alpha(x)) + \alpha _{\mu\nu}(r_H(x), u^\alpha(x), r) + h(r_H(x), u^\alpha(x), r) P_{\mu\nu}(u^\sigma(x)) \Big) dx^\mu dx^\nu \cr
    & - 2 r^{-2} \Big( 1 + j(r_H(x), u^\alpha(x), r) \Big) u_\mu(x) dx^\mu dr \bigg].
\end{align}
Note here we use a different convention on the sign of the scalar perturbation $j$ from \cite{Wu2111}. For the probe scalar field, we have
\begin{align}
  \Phi(x^\mu, r) = \phi(x^\mu) + \sum_{n \geq 1} \Phi^{(n)}(x^\mu, r). \label{eq: global on-shell Phi}
\end{align}
Here the change from $\phi_0$ to $\phi(x^\mu)$ is the same as $r_H$ or $u^\mu$ being promoted to be $x^\mu$-dependent. In order to make the scalar probe on-shell again, we also need to add the perturbations $\sum_{n \geq 1} \Phi^{(n)}(x^\mu, r)$ for the scalar probe.

\subsection{Solving the perturbations to the second order}

We expand \cref{eq: global on-shell metric rAdS} to the first order of boundary derivative and get
\begin{align}\label{eq: 1st-order expanded metric rAdS}
  ds^2 =&\; r^\frac{2p}{p-q} \bigg[ - \bigg( f(r) - \frac{(p+1)r_H^{p}}{r^{p+1}}\delta r_H + k^{(1)}(r) \bigg) dv^2 + 2 \big( (f-1) \delta \beta_i + w^{(1)}_i(r) \big) dvdx^i \cr
  & + (\delta_{ij} + \alpha_{ij}^{(1)}(r) + h^{(1)}(r) \delta_{ij}) dx^idx^j + 2 r^{-2} (1 + j^{(1)}(r)) dvdr - 2 r^{-2} \delta\beta_i dx^idr \bigg],
\end{align}
where $\delta \psi = x^\mu \partial_\mu \psi$ (with $\psi$ stands for either $r_H$, $\beta_i$ or $\phi$ in later texts). Then put the above first-order expanded metric into the traceless-tensor components, the $(ri)$-components, the $(E_{rN} - T_{rN})(dr)^N$ components, the $(rr)$-components of the Einstein equation:
\begin{align}
  E_{ij} - \frac1{p-q} \delta_{ij} \delta^{kl} E_{kl} - \left( T_{ij} - \frac1{p-q} \delta_{ij} \delta^{kl} T_{kl} \right) = 0,
  \label{eq: Einstein eq(ij)} \\
  E_{ri} - T_{ri} = 0, \label{eq: Einstein eq(ri)} \\
  g^{r0} (E_{r0} - T_{r0}) + g^{rr} (E_{rr} - T_{rr}) = 0, \label{eq: Einstein eq(M=r constraint)}\\
  E_{rr}-T_{rr} = 0, \label{eq: Einstein eq(rr)}
\end{align}
and also the EOM of $\varphi$ \cref{eq: EOM varphi}, one gets the differential equations for the first-order metric perturbations
\begin{align}\label{eq: diff eq of 1st order metric pert rAdS}
  & \partial_r (r^{p+2} f(r) \partial_r \alpha^{(1)}_{ij}(r)) + 2p r^{p-1} \sigma_{ij} = 0, \cr
  & \partial_r (r^{p+2} \partial_r w^{(1)}_i(r)) + p r^{p-1} \partial_0 \beta_i = 0, \cr
  & (r^{p+1} k_{(1)})' - 2 (p+1) r^p j_{(1)} + (p-q) \left( r^{p+1} - \frac{p-1}{2p} r_H^{p+1} \right) h'_{(1)} + 2 r^{p-1} \partial \beta = 0, \cr
  & (p-q) r h''_{(1)} + 2(p-q) h'_{(1)} - 2p j'_{(1)} = 0, \cr
  & (r^{p+1} k_{(1)})' - r^{p+1} f j'_{(1)} + 2(p+1) r^p j_{(1)} + \frac{p-q}{2} r^{p+1} f h'_{(1)} + r^{p-1} \partial \beta = 0.
\end{align}
The solutions are
\begin{align}\label{eq: 1st-order metric pert. general form}
  \alpha^{(1)}_{ij}(r) &= F(r) \sigma_{ij}, \qquad w_i^{(1)}(r) = F_w(r) \partial _0\beta_i, \qquad h_{(1)}(r) = F_h(r) \partial \beta, \cr
  j_{(1)}(r) &= F_j(r) \partial \beta, \qquad k_{(1)}(r) = F_k(r) \partial \beta,
\end{align}
where
\begin{align}
  & F = \frac{2}{r} {}_2F_1\left( 1, \frac{1}{p+1}, \frac{p+2}{p+1}, \left( \frac{r_H}{r} \right)^{p+1} \right) + \frac{2 \ln f(r)}{(p+1) r_H}, \ \  F_w = \frac{1}{r}, \ \  F_h = \frac1{p-q} F, \cr
  & F_j = - \frac{1}{p} \frac{ r^p-r_H^p }{ r^{p+1} - r_H^{p+1} } + \frac{1}{2p} F, \qquad F_k = - \frac2{pr} + \frac{1}{p} \left( 1+ \frac{(p-1) r_H^{p+1}}{2 r^{p+1}} \right) F.
\end{align}
Here we write $F$ in hypergeometric series for all the cases of $p \geq 2$ and $1\leq q\leq p-1$, which is different from \cite{Wu2111} where $F$ is given in specific form for each $p$. One can see that the solutions of the first-order metric perturbations are the same as in \cite{Wu2111}, the scalar probe does not change them. This is because by \cref{eq: bulk energy-momentum tensor} we know the scalar probe enters into the Einstein equation via second-order terms, it does not affect the metric perturbations at the first order. The appearance of $\Phi$ in the metric perturbations should start in the second order. This has been explained in \cite{Bhattacharyya0806}, thus we only need to solve the perturbation of $\Phi$ in the first order.

Just like the metric, we expand the scalar probe to the first order as
\begin{align}
  \Phi(x^\mu,r) = \delta \phi + \Phi^{(1)}(r), \label{eq: expansion of global on-shell Phi to 1st order}
\end{align}
here we set $\phi(x^\mu=0) = 0$. Putting the above into the EOM of $\Phi$ \cref{eq: EOM Phi}, one has the differential equation for $\Phi^{(1)}$ as
\begin{align}\label{eq: diff eq of Phi(1)}
  \partial_r (r^{p+2} f(r) \partial_r \Phi^{(1)}(r)) + p r^{p-1} \partial_0 \phi= 0.
\end{align}
Compared with the equation of $\alpha_{ij}^{(1)}$ in \cref{eq: diff eq of 1st order metric pert rAdS}, $\Phi^{(1)}$ should be
\begin{align}
  \Phi^{(1)}(r) = F_\phi(r) \partial_0 \phi = \frac12 F \partial_0 \phi.
\end{align}
With the solved $\Phi^{(1)}$, the global on-shell form of the scalar probe \cref{eq: global on-shell Phi} should now read
\begin{align}\label{eq: global on-shell Phi with 1st order solved}
  \Phi(x^\mu,r) = \phi(x) + F_\phi(r_H(x),r) D\phi(x,r) + \sum_{n \geq 2} \Phi^{(n)}(x^\mu, r).
\end{align}
In the above, $\Phi^{(1)}(r)$ has been solved and the global form of it is
\begin{align}
  \Phi^{(1)}( x,r ) = F_\phi(r_H(x),r) D\phi(x,r),
\end{align}
where $D = u^\mu \partial_\mu$ is the temporal derivative of the relativistic fluid. So the summation of $\Phi^{(n)}$ begins with the second order in \cref{eq: global on-shell Phi with 1st order solved}, which will be used to get the second-order expanded form of $\Phi$ later on.

To solve the second order, the global on-shell metric \cref{eq: global on-shell metric rAdS} needs to be expanded to the second order as
\begin{align}\label{eq: 2nd order expanded metric rAdS}
  ds^2 =&\; r^\frac{2p}{p-q} \bigg[ - \bigg( f - (1-f) \delta\beta_i \delta\beta_i - \frac{(p+1)r_H^{p}}{r^{p+1}}(\delta r_H +\frac{1}{2} \delta^2r_H+\delta r_H^{(1)}) \cr
  & - \frac{p(p+1) r_H^{p-1}}{2r^{p+1}}(\delta r_H)^2 + (F_k+\delta F_k)\partial \beta + F_k(\delta\partial \beta + \delta\beta_i \partial_0\beta_i) + 2 F_w \delta\beta_i \partial_0\beta_i \cr
  & + k^{(2)}(r) \bigg) dv^2 + 2 \bigg( (f-1)(\delta\beta_i + \frac12 \delta^2\beta_i) + F_w (\partial_0\beta_i + \delta\partial_0\beta_i + \delta\beta_j\partial_j\beta_i) \cr
  & - \frac{(p+1)r_H^{p}}{r^{p+1}}\delta r_H \delta\beta_i + F_k \partial \beta \delta\beta_i - F \delta\beta_j \partial_{(i}\beta_{j)} + w_i^{(2)}(r) \bigg) dvdx^i \cr
  & + \bigg( \delta_{ij} + (1-f)\delta\beta_i\delta\beta_j - 2 F_w \delta\beta_{(i}\partial_{|0|}\beta_{j)} + (F + \delta F) \partial_{(i}\beta_{j)} \cr
  & + F \left( \delta\partial_{(i} \beta_{j)} + \delta\beta_{(i} \partial_{|0|}\beta_{j)} \right) + \alpha_{ij}^{(2)}(r) + h^{(2)}(r) \delta_{ij} \bigg) dx^idx^j \cr
  & + 2r^{-2} \bigg( 1 + \frac12 \delta\beta_i\delta\beta_i + (F_j+\delta F_j) \partial \beta + F_j(\delta\partial \beta + \delta\beta_i\partial_0\beta_i) + j^{(2)}(r) \bigg) dvdr \cr
  & - 2 r^{-2} \bigg( \delta\beta_i + \frac12 \delta^2\beta_i + F_j \partial\beta \delta\beta_i \bigg) dx^i dr \bigg],
\end{align}
and the global on-shell $\Phi$ with the solved first-order perturbation \cref{eq: global on-shell Phi with 1st order solved} should be expanded to the second order as
\begin{align}\label{eq: 2nd order expanded Phi}
  \Phi(x^\mu, r) = \delta\phi + \frac12 \delta^2\phi + \big( F_\phi + \delta F_\phi \big) (\partial_0 \phi + \delta\partial_0 \phi + \delta \beta_i \partial_i \phi) + \Phi^{(2)}(r).
\end{align}
In \cref{eq: 2nd order expanded metric rAdS} and \cref{eq: 2nd order expanded Phi}, we have defined
\begin{align}
  \delta \mathcal F(r_H(x),r) = - \frac{\mathcal{F}(r) + r \mathcal{F}'(r)}{r_H} \delta r_H,
\end{align}
where $\mathcal F$ stands for $F$, $F_j$, $F_k$ or $F_\phi$.

The second-order metric perturbations are solved by bringing the second-order expanded metric into \cref{eq: Einstein eq(ij),eq: Einstein eq(M=r constraint),eq: Einstein eq(ri),eq: Einstein eq(rr)} and the EOM of $\varphi$ \cref{eq: EOM varphi}, the results will be omitted in this paper. $\Phi^{(2)}$ are solved by putting \cref{eq: 2nd order expanded Phi} into the EOM of $\Phi$ \cref{eq: EOM Phi} and the solution can be summarized by
\begin{align}\label{eq: solution of Phi(2) rAdS}
  \Phi^{(2)}(r) =& \bigg[ \frac{p-2}{2(p-1)} \frac{1}{r^2} + \bigg( \frac{3-p}{2(p-1)(p+1)} + \frac{1}{(p+1)^2} H_\frac{2}{p+1} \bigg) \frac{1}{r^{p+1}} \bigg] s_{\phi1} \cr
  & + \bigg[ \frac{1}{2(p-1)} \frac{1}{r^2} - \frac{1}{(p-1)(p+1)} \frac{1}{r^{p+1}} \bigg] s_{\phi2} \cr
  & - \bigg( \frac{3}{2p(p+1)} - \frac{1}{p(p+1)^2} H_\frac{2}{p+1} \bigg) \frac{1}{r^{p+1}} \mathfrak S_{\phi3} \cr
  & + \bigg( \frac{1}{2(p+1)} + \frac{1}{(p+1)^2} H_\frac{2}{p+1} \bigg) \frac{1}{r^{p+1}} \mathfrak S_{\phi4}.
\end{align}
One can see that the solutions of $\Phi^{(n)}(r)$ are independent of the number of compactified directions $q$. Furthermore, one may infer that the final results of the transport coefficients related to both the external scalar field and the driving force exerted on the fluid are $q$-independent.

There are still two constraints built from the Einstein equation $(E_{\mu N} - T_{\mu N}) (dr)^N = 0$ with $\mu$ takes 0 or $i$. For the first order, $(E_{\mu N} - T_{\mu N}) (dr)^N = 0$ gives
\begin{align}\label{eq: Navier-Stokes rAdS 1st order}
   \frac1{r_H} \partial_0 r_H = - \frac1{p} \partial \beta, \qquad  \frac{1}{r_H} \partial _i r_H = - \partial _0 \beta_i.
\end{align}
Note $(dr)^N = g^{NP} (dr)_P = g^{NP} \delta_P^r = g^{Nr}$. In the second order, one gets
\begin{align}
  \partial_0 r_H^{(1)} =&\; \frac{2q}{p^2(p+1)(p-q)} \mathfrak S_3 + \frac{2}{p(p+1)} \mathfrak S_5 + \frac{1}{p(p+1)}\mathfrak S_{\phi 1},
  \label{eq: Navier-Stokes(0) rAdS 2nd order} \\
  \partial_i r_H^{(1)} =&\; \frac{2q~ \mathbf v_4 + 2(p-q)(p-1) \mathbf v_5}{p(p+1)(p-q-1)}+ \frac{(p-1)(p-2q+2) - pq}{p(p-q)(p+1)} \mathfrak V_1 \cr
  & - \frac{p+2}{p(p+1)} \mathfrak V_2 - \frac{2p^2 - 3p + 2}{p(p+1)} \mathfrak V_3 - \frac{1}{p+1}\mathfrak V_{\phi}. \label{eq: Navier-Stokes(i) rAdS 2nd order}
\end{align}
Note there is a factor $(p-q-1)$ appearing in \cref{eq: Navier-Stokes(i) rAdS 2nd order}, indicating its failure at $p-q=1$. What's more,  $\mathfrak S_5$ and $\mathfrak V_{2,3}$ do not exist when the spatial dimension of the fluid is 1. So \cref{eq: Navier-Stokes(0) rAdS 2nd order,eq: Navier-Stokes(i) rAdS 2nd order} are valid for $p \geq 2$ and $1 \leq q \leq p-2$. When $p-q=1$, we can use the identity that $q \mathbf v_4 + (p-1)(p-q) \mathbf v_5 = (p-1) (p-q-1) \mathbf v_3$ to reexpress the Navier-Stokes equation as
\begin{align}\label{eq: Navier-Stokes rAdS 2nd order p-q=1}
  \partial_0 r_H^{(1)} &= \frac{2(p-1)}{p^2 (p+1)} \mathfrak S_3 + \frac{1}{p(p+1)}\mathfrak S_{\phi 1}, \cr
  \partial_i r_H^{(1)} &= \frac{2(p-1)}{p (p+1)} \mathbf v_3 - \frac{2 (p-1) (p-2)}{p (p+1)} \mathfrak V_1 - \frac{1}{p+1}\mathfrak V_{\phi}.
\end{align}
With the present form, \cref{eq: Navier-Stokes rAdS 2nd order p-q=1} is valid for $p \geq 2$ and $q=p-1$. These are the cases that the spatial dimension of the fluid is 1.

The physical means of the equations \cref{eq: Navier-Stokes rAdS 1st order} and \cref{eq: Navier-Stokes(0) rAdS 2nd order,eq: Navier-Stokes(i) rAdS 2nd order,eq: Navier-Stokes rAdS 2nd order p-q=1} derived from $(E_{\mu N} - T_{\mu N}) (dr)^N = 0$ are the Navier-Stokes equations separately at the first and second orders \cite{Bhattacharyya0712}.
We can see from the second-order Navier-Stokes equations \cref{eq: Navier-Stokes(0) rAdS 2nd order,eq: Navier-Stokes(i) rAdS 2nd order,eq: Navier-Stokes rAdS 2nd order p-q=1} that there indeed are contributions from the external scalar field, whose spatial viscous tensors are $\mathfrak S_{\phi 1}$ and $\mathfrak V_\phi$. The definitions of all the spatial viscous tensors can be found in \cref{tab: 2nd order spatial viscous terms}.
\begin{table}[h]
\centering
\begin{tabular}{|l|l|l|}
  \hline
  Scalars of $\mathrm{SO}(d-1)$                                 &                 Vectors of $\mathrm{SO}(d-1)$                      &\quad  Tensors of $\mathrm{SO}(d-1)$ \\ \hline\hline
 $\mathbf{s}_1=\frac1{r_H}\partial_0^2r_H$               & $\mathbf{v}_{1i} = \frac1{r_H} \partial_0\partial_i r_H$ & $\mathbf{t}_{1ij} = \frac1{r_H} \partial_i\partial_j r_H - \frac1{d-1} \delta_{ij} \mathbf{s}_3$ \\
  $\mathbf{s}_2 = \partial_0\partial_i\beta_i$                & $\mathbf{v}_{2i} = \partial_0^2\beta_i$       & $\mathbf{t}_{2ij} = \partial_0 \Omega_{ij}$ \\
  $\mathbf{s}_3 = \frac1{r_H}\partial_i^2r_H$             & $\mathbf{v}_{3i} = \partial_j^2\beta_i$        & $\mathbf{t}_{3ij} = \partial_0\sigma_{ij}$ \\
  $\mathfrak S_1 = \partial_0\beta_i\partial_0\beta_i$   & $\mathbf{v}_{4i}=\partial_j\Omega_{ij}$      & $\mathfrak T_{1ij} = \partial_0\beta_i\partial_0\beta_j - \frac1{d-1} \delta_{ij} \mathfrak S_1$ \\
  $\mathfrak S_2 = \epsilon_{ijk} \partial_0 \beta_i \partial_j \beta_k$  & $\mathbf{v}_{5i} = \partial_j\sigma_{ij}$  & $\mathfrak T_{2ij} = \sigma_{[i}^{~~k} \Omega_{j]k}$ \\
  $\mathfrak S_3 = (\partial_i\beta_i)^2$                       & $\mathfrak V_{1i} = \partial_0\beta_i\partial\beta$     & $\mathfrak T_{3ij} = \Omega_{ij} \partial \beta$ \\
  $\mathfrak S_4 = \Omega_{ij} \Omega_{ij}$              &  $\mathfrak V_{2i} = \partial_0\beta_j \Omega_{ij}$  & $\mathfrak T_{4ij}=\sigma_{ij}\partial\beta$ \\
  $\mathfrak S_5=\sigma_{ij}\sigma_{ij}$                      & $\mathfrak V_{3i} = \partial_0\beta_j \sigma_{ij}$  & $\mathfrak T_{5ij} = \Omega_i^{~k}\Omega_{jk} - \frac1{d-1} \delta_{ij} \mathfrak S_4$ \\
  $\mathbf{s}_{\phi1} = \partial_0^2 \phi$                  &  $\mathfrak V_{\phi i} = \partial_0 \phi \partial_i \phi$  &  $\mathfrak T_{6ij} = \sigma_i^{~k}\sigma_{jk} - \frac1{d-1} \delta_{ij} \mathfrak S_5$ \\
  $\mathbf{s}_{\phi2} = \partial_i^2 \phi$                   &                                                             & $\mathfrak T_{7ij} = \sigma_{(i}^{~~k} \Omega_{j)k}$ \\
  $\mathfrak S_{\phi1} = (\partial_0 \phi)^2$               &                                                             &  $\mathfrak T_{\phi ij} = \partial_i \phi \partial_j \phi - \frac{1}{d-1} \delta_{ij} (\partial \phi)^2$ \\
  $\mathfrak S_{\phi2} = (\partial_i \phi)^2$                &                                                             & \\
  $\mathfrak S_{\phi3} = \partial \beta \partial_0 \phi$ &                                                             & \\
  $\mathfrak S_{\phi4} = \partial_0 \beta_i \partial_i \phi$ &                                                            & \\
\hline
\end{tabular}
\caption{\label{tab: 2nd order spatial viscous terms} Here $d-1$ is the spatial dimensions of the fluid, for the cases of reduced AdS black hole and compactified Dp-brane, $d-1=p-q$. While for the reduced Dp- and smeared Dp-brane, $d-1=p$. In this table, $\mathbf{s}_{\phi1,2}$, $\mathfrak S_{\phi1,2,3,4}$, $\mathfrak V_{\phi i}$ and $\mathfrak T_{\phi ij}$ are the newly appearing spatial viscous terms defined for the external scalar field $\phi$.}
\end{table}

\subsection{The stress-energy tensor and driving force of the fluid}\label{subsec: deriving the force}

The stress-energy tensor of the dual fluid can be extracted by taking the boundary limit of the surface tensor in the bulk \cref{eq: surface tensor in the bulk general form} with $c_{1,2}$, $\gamma$ and $e^{\varphi}$ taking their values in \cref{eq: action in CR form rAdS}:
\begin{align}\label{eq: surface tensor rAdS}
  T_{MN} =&\; \frac{1}{\kappa_{p-q+2}^2} \lim_{r\to\infty} \left( \frac rL \right)^\frac{p(p-q-1)}{p-q} \bigg[ K_{MN} - h_{MN}K - \frac{p}{L} \left( \frac rL \right)^{-\frac{q}{p-q}} h_{MN} \cr
 &  - \frac{L}{2(p-1)} \left( \frac rL \right)^\frac{q}{p-q} \left( \bar{\nabla}_M \Phi \bar{\nabla}_N \Phi - \frac12 h_{MN} \left( \bar{\nabla} \Phi \right)^2 \right) \bigg].
\end{align}
Substituting the second-order expanded metric of the reduced AdS black hole \cref{eq: 2nd order expanded metric rAdS} and the second-order solution of the scalar probe \cref{eq: solution of Phi(2) rAdS} into the above, one has the stress-energy tensor of the boundary fluid (The $T_{Mr}$ components are all zero)
\begin{align}\label{eq: stress-energy tensor rAdS}
  T_{\mu\nu} =&\; \frac{1}{2 \kappa_{p-q+2}^2} \Bigg\{ {r_H^{p+1} \over L^{p+2}} \left( p~ u_\mu u_\nu + P_{\mu\nu} \right) - \left( \frac{r_H}{L} \right) ^p \bigg( 2\sigma_{\mu\nu} + \frac{2q}{p(p-q)} P_{\mu\nu} \partial u \bigg) \cr
  & + \frac{r_H^{p-1}}{L^{p-2}} \Bigg[ \bigg( \frac12 + \frac{1}{p+1} H_\frac{2}{p+1} \bigg)\cdot 2\bigg( \sideset{_\langle}{}{\mathop D}\sigma_{\mu\nu\rangle} + \frac1{p-q} \sigma_{\mu\nu} \partial u \bigg) \cr
  & + \bigg( \frac{3q}{2p} - \frac{q}{p(p+1)} H_\frac{2}{p+1} \bigg) \frac{2\sigma_{\mu\nu} \partial u}{p-q} + \frac{1}{2} \cdot 4\sigma_{\langle\mu}^{~~\rho}\sigma_{\nu\rangle\rho} \cr
  & + \bigg( - 1 + \frac{2}{p+1} H_\frac{2}{p+1} \bigg) \cdot 2 \sigma_{\langle\mu}^{~~\rho} \Omega_{\nu\rangle\rho} - \frac1{p-1} \partial_{\langle \mu} \phi \partial_{\nu \rangle} \phi \Bigg] \cr
  & + \frac{r_H^{p-1}}{L^{p-2}} P_{\mu\nu} \Bigg[ \bigg( \frac{q}{p(p-q)} + \frac{2q}{p(p+1)(p-q)} H_\frac{2}{p+1} \bigg) D(\partial u) \cr
  & + \Bigg( \frac{q(3q-p)}{p^2(p-q)^2} + \frac{2q}{p^2(p+1)(p-q)} H_\frac{2}{p+1} \Bigg) (\partial u)^2 + \frac{q}{2p(p-q)} \cdot 4 \sigma_{\alpha\beta}^2 \cr
  & - \frac{q}{p(p-1)(p-q)} (\partial_{\bot} \phi)^2 \Bigg] \Bigg\}.
\end{align}
Here $(\partial_{\bot} \phi)^2 = P^{\rho\sigma} \partial_\rho \phi \partial_\sigma \phi$, and we define the spatial-projected symmetric traceless tensor
\begin{align}
  T_{\langle \mu\nu \rangle} = P_\mu^\rho P_\nu^\sigma T_{(\rho\sigma)} - \frac1{d-1} P_{\mu\nu} P^{\rho\sigma} T_{\rho\sigma}.
\end{align}
One can see that the second-order stress tensor contains the contributions from the external scalar field $\phi$. If we define the viscous terms of $\phi$ in the stress tensor as
\begin{align}
  T^\phi_{\mu\nu} = \lambda_\phi \partial_{\langle \mu} \phi \partial_{\nu \rangle} \phi + \xi_\phi P_{\mu\nu} (\partial_{\bot} \phi)^2.
\end{align}
Then we can read the transport coefficients related to $\phi$ as
\begin{align}
  \lambda_{\phi} = \frac1{2\kappa_{p-q+2}^2} \frac{-1}{p-1} \frac{r_H^{p-1}}{L^{p-2}}, \qquad
  \xi_{\phi} = \frac1{2\kappa_{p-q+2}^2} \frac{-q}{p(p-1)(p-q)} \frac{r_H^{p-1}}{L^{p-2}},
\end{align}
which satisfies the identity that $\xi_\phi = \frac12 \gamma^2 \lambda_\phi$.

We define the covariant derivative on a hypersurface at $r = \text{const.}$ as $\bar \nabla_M = h_M^N \nabla_N$, and note the $\nabla_N$ here is the bulk covariant derivative restricted on the hypersurface. Then we will have $\bar \nabla^M h_{NP} = 0$ and $\bar \nabla^M \varphi(r) = 0$. The former is because $h_{MN}$ is compatible with the covariant derivative on the hypersurface. The latter is due to any function of $r$ is also a constant on a hypersurface at constant $r$. Now, let $\bar \nabla^M$ act on $T_{MN}$ in \cref{eq: surface tensor rAdS} one has
\begin{align}
  \bar \nabla^M T_{MN} = \frac{1}{\kappa_{p-q+2}^2} \lim_{r\to\infty} \left( \frac rL \right)^\frac{p(p-q-1)}{p-q} \bigg[ \bar \nabla^M K_{MN} - \bar \nabla_N K - \frac{L}{2(p-1)} \left( \frac rL \right)^\frac{q}{p-q} \bar \nabla^2 \Phi \bar \nabla_N \Phi \bigg]
\end{align}
The Codazzi's equation $\bar \nabla^M K_{MN} - \bar \nabla_N K = - h_N^P n^Q R_{PQ}$ will help here.
With the Einstein equation \cref{eq: Einstein equation}:
\begin{align}
  R_{MN} - \frac12 g_{MN}R =&\; \frac12 \Big(\partial_M \varphi \partial_N \varphi - \frac12 g_{MN} (\partial \varphi)^2\Big) - \frac12 g_{MN} V(\varphi) \cr
                               & + \frac12 \Big(\partial_M \Phi \partial_N \Phi - \frac12 g_{MN} (\partial \Phi)^2 \Big),
\end{align}
and given that $h_N^P n^Q g_{PQ} = h^P_N n_P = 0$ and $h_N^P \nabla_P \varphi = \bar \nabla_N \varphi = 0$, one then has $h_N^P n^Q R_{PQ} = \frac12 \nabla_n \Phi \bar \nabla_N \Phi$. $\nabla_n = n^M \nabla_M$ is the directional derivative along the unit normal vector $n^M$. Then we arrive at
\begin{align}
  \bar \nabla^M T_{MN} &= \frac{1}{2 \kappa_{p-q+2}^2} \lim_{r\to\infty} \bigg[ \left( \frac rL \right)^\frac{p(p-q-1)}{p-q} \bigg( - \nabla_n \Phi - \frac{L}{p-1} \left( \frac rL \right)^\frac{q}{p-q} \bar \nabla^2 \Phi \bigg) \bar\nabla_N \Phi \bigg].
\end{align}
By the fact that $T_{Mr}=0$, the nontrivial components left from the above gives
\begin{align}
  \bar\nabla^\mu T_{\mu\nu} &= \frac{1}{2 \kappa_{p-q+2}^2} \bigg[ \lim_{r\to\infty} \left( \frac rL \right)^\frac{p(p-q-1)}{p-q} \bigg( - \nabla_n \Phi - \frac{L}{p-1} \left( \frac rL \right)^\frac{q}{p-q} \bar \nabla^2 \Phi \bigg) \bigg] \bigg[ \lim_{r\to\infty} \bar\nabla_\nu \Phi \bigg] \cr
  &= \frac{1}{2 \kappa_{p-q+2}^2} \bigg[ \lim_{r\to\infty} \left( \frac rL \right)^\frac{p(p-q-1)}{p-q} \bigg( - \nabla_n \Phi - \frac{L}{p-1} \left( \frac rL \right)^\frac{q}{p-q} \bar \nabla^2 \Phi \bigg) \bigg] \left( \frac rL \right)^\frac{2p}{p-q} \bar\nabla_\nu \phi \cr
  &= \frac{1}{2 \kappa_{p-q+2}^2} \bigg[ \lim_{r\to\infty} \left( \frac rL \right)^\frac{p(p-q+1)}{p-q} \bigg( - \nabla_n \Phi - \frac{L}{p-1} \left( \frac rL \right)^\frac{q}{p-q} \bar \nabla^2 \Phi \bigg) \bigg] \bar\nabla_\nu \phi.
\end{align}
Here we use the near-boundary behavior of the bulk scalar probe
\begin{align}
  \lim_{r\to\infty} \bar\nabla_\nu \Phi = C(r) \bar\nabla_\nu \phi,
\end{align}
where $C(r)$ is the conformal factors in the metric of Chamblin-Reall gravities \cref{eq: background solution rAdS,eq: background solution cDp,eq: background solution sDp}, as listed in \cref{tab: powers of the conformal factors}. Comparing with the forced Navier-Stokes equation of the fluid \cref{eq: the forced NS eq}, we finally obtain the holographic definition for the Lagrangian density of the fluid:
\begin{align}\label{eq: Lagrangian definition rAdS}
  e^{-\phi} \mathcal L = \frac{1}{2 \kappa_{p-q+2}^2} \lim_{r\to\infty} \left( \frac rL \right)^\frac{p(p-q+1)}{p-q} \bigg[ - \nabla_n \Phi - \frac{L}{p-1} \left( \frac rL \right)^\frac{q}{p-q} \bar \nabla^2 \Phi \bigg].
\end{align}
This definition can be used to calculate the Lagrangian density. When $q=0$, \cref{eq: Lagrangian definition rAdS} goes back to the case for the conformal AdS black hole of general dimensions \cite{Ashok1309}.
\begin{table}[h!]
\centering
\begin{tabular}{|c|c|c|c|}
  \hline
               & \makecell[c]{reduced AdS \\ black hole} & \makecell[c]{reduced compactified \\ Dp-brane} & \makecell[c]{reduced smeared \\ Dp-brane} \\ \hline
    $C(r)$    & $\left( \frac rL \right)^\frac{2p}{p-q}$ & $\left( \frac rL \right)^\frac{9-p}{p-q}$ & $\left( \frac rL \right)^\frac{9-p-q}{p}$ \\
  \hline
\end{tabular}
\caption{\label{tab: powers of the conformal factors} The conformal factors in the metrics of the Chamblin-Reall gravity with one background scalar field.}
\end{table}

Using the second-order on-shell metric and the solution of $\Phi$, we can derive the Lagrangian density as
\begin{align}\label{eq: Lagrangian result rAdS}
  e^{-\phi} \mathcal{L} =&\; \frac{1}{2 \kappa_{p-q+2}^2}\Bigg\{ -\left( \frac{r_H}{L} \right) ^p D \phi + \frac{r_H^{p-1}}{L^{p-2}} \Bigg[ \bigg( \frac{3-p}{2(p-1)} + \frac{1}{p+1} H_\frac{2}{p+1} \bigg) u^\mu u^\nu \partial_\mu \partial_\nu \phi \cr
  & - \frac1{p-1} \partial^2_{\bot} \phi + \bigg( -\frac{3}{2p} + \frac{1}{p(p+1)} H_\frac{2}{p+1} \bigg) \partial u D \phi \cr
  & + \bigg( \frac12 + \frac{1}{p+1} H_\frac{2}{p+1} \bigg) D u^{\mu} \partial_{\mu} \phi \Bigg] \Bigg\}.
\end{align}
The Lagrangian density in relativistic hydrodynamics is defined by
\begin{align}\label{eq: Lagrangian definition in hydrodynamics}
  e^{-\phi} \mathcal{L} = - \zeta_\phi D\phi + \xi_{\phi1} u^\mu u^\nu \partial_\mu \partial_\nu \phi + \xi_{\phi2}  \partial^2_{\bot} \phi + \xi_{\phi3}  \partial u D \phi + \xi_{\phi4}  D u^{\mu} \partial_{\mu} \phi.
\end{align}
Here $\zeta_\phi$ and $\xi_{\phi1,\phi2,\phi3,\phi4}$ are respectively the first and second order coefficients of the Lagrangian density or the driving force. One does not count the order of $\nabla^\mu \phi$ in the definition of the driving force $f^\mu = e^{-\phi} \mathcal{L} \nabla^\mu \phi$. We can read all the transport coefficients of the driving force for the relativistic fluid by comparing \cref{eq: Lagrangian result rAdS} with \cref{eq: Lagrangian definition in hydrodynamics}
\begin{align}
  \zeta_\phi  &= \frac1{2\kappa_{p-q+2}^2} \frac{r_H^p}{L^p}, \hspace{22mm} \xi_{\phi1} = \frac1{2\kappa_{p-q+2}^2} \left[ \frac{3-p}{2(p-1)} + \frac{1}{p+1} H_\frac{2}{p+1} \right] \frac{r_H^{p-1}}{L^{p-2}}, \cr
   \quad \xi_{\phi2} &= \frac1{2\kappa_{p-q+2}^2} \frac{-1}{p-1} \frac{r_H^{p-1}}{L^{p-2}}, \qquad \xi_{\phi3} = \frac1{2\kappa_{p-q+2}^2} \left[- \frac{3}{2p} + \frac{1}{p(p+1)} H_\frac{2}{p+1} \right] \frac{r_H^{p-1}}{L^{p-2}}, \cr
   \xi_{\phi4} &= \frac1{2\kappa_{p-q+2}^2} \left[ \frac12 + \frac{1}{p+1} H_\frac{2}{p+1} \right] \frac{r_H^{p-1}}{L^{p-2}}.
\end{align}
We can see that all the coefficients of the driving force have nothing to do with the number of compactified directions $q$ of the AdS black hole. This means the above Lagrangian coefficients should coincide only numerically with those of \cite{Bhattacharyya0806} with $p=3$ and those of \cite{Ashok1309} with general $p$. The dimensions of these coefficients are not the same unless one sets $q=0$ here. One can check that if the gradient terms of $\phi$ in the Lagrangian in \cite{Bhattacharyya0806,Ashok1309} are written in the forms here, the numerical values of the Lagrangian coefficients are indeed the same.

Another interesting observation on the four second-order coefficients of the driving force is that only two of them are independent since there are two identities existing among $\xi_{\phi1,\phi2,\phi3,\phi4}$ as
\begin{align}
  \xi_{\phi3} = c_s^2 \xi_{\phi1} + \xi_{\phi_2}, \qquad \qquad \xi_{\phi4} = \xi_{\phi1} - (p-2) \xi_{\phi_2}.
\end{align}
Where $c_s^2 = \frac{1}{p}$ is the sound speed of the dual fluid \cite{Wu2111}. Besides, one can find another four apparent relations between the coefficients of the driving force and those of the stress tensor:
\begin{align}
  \zeta_\phi = \eta, \qquad \xi_{\phi2} = \lambda_\phi, \qquad \xi_{\phi3} = - \frac{2c_s^2}{(p-q) \gamma^2} \eta \tau_\pi^*, \qquad \xi_{\phi4} = \eta \tau_\pi.
\end{align}

\section{The reduced Dp-brane and the reduced compactified Dp-brane}

\subsection{The background and scalar probe}

The full action for the reduced compactified Dp-brane in Chamblin-Reall form \cite{Wu2111} with the manually added scalar probe can be written as
\begin{align}\label{eq: action in CR form cDp}
  S =&\, \frac1{2\kappa_{p-q+2}^2} \int d^{p-q+2}x \sqrt{-g} \left[ R - \frac12 (\partial \varphi)^2 - V(\varphi) - \frac12 (\partial \Phi)^2 \right] \cr
  & - \frac1{\kappa_{p-q+2}^2} \int d^{p-q+1}x \sqrt{-h} \left[ K + \frac{9-p}{2L} \, e^{- \frac{\gamma}{2} \varphi} - \frac{L}{8} e^{\frac\gamma2 \varphi} (\bar \partial \Phi)^2 \right],
\end{align}
where the potential and the parameter $\gamma$ are
\begin{align}
  V(\varphi) = - \frac{(7-p)(9-p)}{2L^2} e^{- \gamma \varphi}, \qquad \gamma^2 = \frac{2(p-3)^2 + 2q(5-p)}{(p-q) (9-p)}.
\end{align}
When $q=0$, \cref{eq: action in CR form cDp} becomes the action of the reduced Dp-brane \cite{Wu1807}. To include the conditions in \cite{Wu1807}, the ranges of $p$ and $q$ in this paper are $1\leq p\leq 4$ and $0\leq q\leq p-1$. From now on, we will refer to the reduced Dp-brane and the reduced compactified Dp-brane together as simply the reduced compactified Dp-brane. All the results of the uncompactified cases can be recovered by setting $q=0$ in the compactified cases. The difference between \cref{eq: action in CR form cDp} from \cref{eq: action in CR form rAdS} is just the numerical factors in the counter terms.

The EOMs that derived from the action of the reduced compactified Dp-brane \cref{eq: action in CR form cDp} have the same form as \cref{eq: Einstein equation,eq: EOM varphi,eq: EOM Phi}, but now
\begin{align}
  V'(\varphi) = \frac{(7-p)(9-p)}{2L^2} \gamma e^{- \gamma \varphi}.
\end{align}
The solution of the EOMs now takes the form
\begin{align}\label{eq: background solution cDp}
  ds^2 =& \left( \frac r{L} \right)^\frac{9-p}{p-q} \left( -f(r) dt^2 + d \vv x^2 + \frac{L^{7-p} dr^2}{r^{7-p} f(r)} \right), \quad e^\varphi = \left( \frac{r}{L} \right)^{\frac{9-p}{2} \gamma}, \quad \Phi = \phi_0.
\end{align}
The emblackening factor here is $f(r) = 1 - r_H^{7-p}/r^{7-p}$.

The global on-shell metric for the reduced compactified Dp-brane is set to be
\begin{align}\label{eq: global on-shell metric cDp}
    ds^2 =&\; r^\frac{9-p}{p-q} \bigg[ - \Big( f(r_H(x),r) + k(r_H(x), u^\alpha(x), r) \Big) u_\mu(x) u_\nu(x) dx^\mu dx^\nu \cr
    & - 2 P_\mu^\rho(u^\alpha(x)) w_\rho(u^\alpha(x), r) u_\nu(x) dx^\mu dx^\nu  \cr
    & + \Big( P_{\mu\nu}(u^\alpha(x)) + \alpha _{\mu\nu}(r_H(x), u^\alpha(x), r) + h(r_H(x), u^\alpha(x), r) P_{\mu\nu}(u^\sigma(x)) \Big) dx^\mu dx^\nu \cr
    & - 2 r^{-\frac{7-p}{2}} \Big( 1 + j(r_H(x), u^\alpha(x), r) \Big) u_\mu(x) dx^\mu dr \bigg].
\end{align}
Compared with \cite{Wu2012}, the scalar and vector perturbations $k$ and $w_\mu$ have the sign flipped. The global on-shell form of $\Phi$ is the same as the case of the reduced AdS black hole \cref{eq: global on-shell Phi}, so we will not offer it again.

\subsection{Solving the perturbations to the second order}

The global on-shell metric of the reduced compactified Dp-brane is expanded to the first order as
\begin{align}\label{eq: 1st-order expanded metric cDp}
  ds^2 =&\; r^\frac{9-p}{p-q} \bigg[ - \bigg( f(r) - \frac{(7-p) r_H^{6-p}}{r^{7-p}}\delta r_H + k^{(1)}(r) \bigg) dv^2 \cr
  & + 2 \big( (f-1) \delta \beta_i + w^{(1)}_i(r) \big) dvdx^i + (\delta_{ij} + \alpha_{ij}^{(1)}(r) + h^{(1)}(r) \delta_{ij}) dx^idx^j \cr
  & + 2 r^{-\frac{7-p}{2}} (1 + j^{(1)}(r)) dvdr - 2 r^{-\frac{7-p}{2}} \delta\beta_i dx^idr \bigg],
\end{align}
Put the above into the different components of the Einstein equation \cref{eq: Einstein eq(ij),eq: Einstein eq(ri),eq: Einstein eq(M=r constraint),eq: Einstein eq(rr)} and the EOM of $\varphi$ \cref{eq: EOM varphi} in the present case, we get the differential equations for all the metric perturbations as
\begin{align}\label{eq: diff eq of 1st order metric pert cDp}
  & \partial_r (r^{8-p} f(r) \partial_r \alpha^{(1)}_{ij}(r)) + (9-p) r^\frac{7-p}{2} \sigma_{ij} = 0, \cr
  & \partial_r (r^{8-p} \partial_r w^{(1)}_i(r)) + \frac{9-p}{2} r^\frac{7-p}{2} \partial_0 \beta_i = 0, \cr
  & (r^{7-p} k_{(1)})' - 2 (7-p) r^{6-p} j_{(1)} + (p-q) \left( r^{7-p} - \frac2{9-p} r_H^{7-p} \right) h'_{(1)} + 2 r^\frac{7-p}{2} \partial \beta = 0, \cr
  & (p-q) r h''_{(1)} + \frac12 (7-p) (p-q) h'_{(1)} - (9-p) j'_{(1)} = 0, \cr
  & (r^{7-p} k_{(1)})' - r^{7-p} f j'_{(1)} - 2(7-p) r^{6-p} j_{(1)} + \frac12 (p-q) r^{7-p} f h'_{(1)} + r^\frac{7-p}{2} \partial \beta = 0.
\end{align}
The first-order metric perturbations share the same form as the case of reduced AdS black hole \cref{eq: 1st-order metric pert. general form}, but the functions have changed to
\begin{align}\label{eq: F functions cDp}
  & F = \frac{4}{(5-p)r^\frac{5-p}{2}} {}_2F_1\left( 1, \frac{5-p}{2(7-p)}, 1+\frac{5-p}{2(7-p)}, \left( \frac{r_H}{r} \right)^{7-p} \right) + \frac{2 \ln f(r)}{(7-p) r_H^\frac{5-p}{2}}, \cr
  & F_w = \frac{2}{(5-p) r^\frac{5-p}{2}}, \quad  F_h = \frac1{p-q} F, \quad F_j = - \frac{2}{9-p} \frac{ r^\frac{9-p}{2} - r_H^\frac{9-p}{2} }{ r^{7-p} - r_H^{7-p} } + \frac{5-p}{2(9-p)} F, \cr
  & F_k = - \frac{4}{(9-p) r^\frac{5-p}{2}} + \frac{1}{9-p} \left( 5-p + \frac{2 r_H^{7-p}}{r^{7-p}} \right) F.
\end{align}
The solutions above cover all the situations of the first-order metric perturbations for the Dp- \cite{Wu1807} and compactified Dp-brane \cite{Wu1508,Wu1604,Wu2012}.

The expansion of the bulk probe scalar $\Phi$ to the first order is the same as \cref{eq: expansion of global on-shell Phi to 1st order}, the differential equation for $\Phi^{(1)}$ is
\begin{align}
  \partial_r (r^{8-p} f(r) \partial_r \Phi^{(1)}(r)) + \frac{9-p}{2} r^\frac{7-p}{2} \partial_0 \phi = 0.
\end{align}
The solution is $\Phi^{(1)} = F_\phi(r) \partial_0 \phi$, where $F_\phi = \frac12 F$ with $F$ takes the result in \cref{eq: F functions cDp}.

The second-order expanded form of \cref{eq: global on-shell metric cDp} is
\begin{align}\label{eq: 2nd order expanded metric cDp}
  ds^2 =&\; r^\frac{9-p}{p-q} \bigg[ - \bigg( f - (1-f) \delta\beta_i \delta\beta_i - \frac{(7-p)r_H^{6-p}}{r^{7-p}}(\delta r_H +\frac{1}{2} \delta^2r_H+\delta r_H^{(1)}) \cr
  & - \frac{(7-p)(6-p) r_H^{5-p}}{2r^{7-p}}(\delta r_H)^2 + (F_k+\delta F_k)\partial \beta + F_k(\delta\partial \beta + \delta\beta_i \partial_0\beta_i) + 2 F_w \delta\beta_i \partial_0\beta_i \cr
  & + k^{(2)}(r) \bigg) dv^2 + 2 \bigg( (f-1)(\delta\beta_i + \frac12 \delta^2\beta_i) + F_w (\partial_0\beta_i + \delta\partial_0\beta_i + \delta\beta_j\partial_j\beta_i) \cr
  & - \frac{(7-p)r_H^{6-p}}{r^{7-p}}\delta r_H \delta\beta_i + F_k \partial \beta \delta\beta_i - F \delta\beta_j \partial_{(i}\beta_{j)} + w_i^{(2)}(r) \bigg) dvdx^i \cr
  & + \bigg( \delta_{ij} + (1-f)\delta\beta_i\delta\beta_j - 2 F_w \delta\beta_{(i}\partial_{|0|}\beta_{j)} + (F + \delta F) \partial_{(i}\beta_{j)} \cr
  & + F \left( \delta\partial_{(i} \beta_{j)} + \delta\beta_{(i} \partial_{|0|}\beta_{j)} \right) + \alpha_{ij}^{(2)}(r) + h^{(2)}(r) \delta_{ij} \bigg) dx^idx^j \cr
  & + 2r^{- \frac{7-p}{2}} \bigg( 1 + \frac12 \delta\beta_i\delta\beta_i + (F_j+\delta F_j) \partial \beta + F_j(\delta\partial \beta + \delta\beta_i\partial_0\beta_i) + j^{(2)}(r) \bigg) dvdr \cr
  & - 2 r^{- \frac{7-p}{2}} \bigg( \delta\beta_i + \frac12 \delta^2\beta_i + F_j \partial\beta \delta\beta_i \bigg) dx^i dr \bigg].
\end{align}
The second-order expanded form of the bulk scalar probe is the same as \cref{eq: 2nd order expanded Phi}. $\delta \mathcal F$ here is defined as
\begin{align}
  \delta \mathcal F(r_H(x),r) = - \frac{(5-p) \mathcal{F}(r) + 2r \mathcal{F}'(r)}{2r_H} \delta r_H,
\end{align}
with $\mathcal F$ still stands for $F$, $F_j$, $F_k$ or $F_\phi$.

The second-order metric perturbations and $\Phi^{(2)}$ can be solved similarly as the reduced AdS black hole, here we still just offer the solution of the scalar probe:
\begin{align}\label{eq: solution of Phi(2) cDp}
  \Phi^{(2)}(r) =& \bigg[ \frac{p-1}{2(5-p)^2} \frac{1}{r^{5-p}} +\bigg( \frac{3-p}{2(5-p)(7-p)} +\frac{1}{(7-p)^2} H_{\frac{5-p}{7-p}} \bigg) \frac{1}{r^{7-p}} \bigg] \mathbf{s}_{\phi 1} \cr
  & + \bigg[ \frac{1}{2(5-p)} \frac{1}{r^{5-p}} - \frac{1}{2(7-p)} \frac{1}{r^{7-p}} \bigg] \mathbf{s}_{\phi 2} \cr
  & - \bigg( \frac{3}{(7-p)(9-p)} - \frac{5-p}{(7-p)^2 (9-p)}H_{\frac{5-p}{7-p}} \bigg) \frac{1}{r^{7-p}} \mathfrak{S}_{\phi 3} \cr
  & + \bigg( \frac{1}{(5-p)(7-p)} + \frac{1}{(7-p)^2}H_{\frac{5-p}{7-p}} \bigg) \frac{1}{r^{7-p}} \mathfrak{S}_{\phi 4}.
\end{align}

The equation $(E_{\mu N} - T_{\mu N}) (dr)^N = 0$ now at first order gives
\begin{align}
  \frac1{r_H} \partial_0 r_H = - \frac2{9-p} \partial \beta, \qquad \frac{1}{r_H} \partial _i r_H = - \frac{2}{5-p} \partial _0 \beta_i.
\end{align}
At the second order, when $p-q \neq 1$, the Navier-Stokes equation is
\begin{align}
  \frac{1}{r_H^{(p-3)/2}} \partial_0 r_H^{(1)} =&\; \frac{4(p-3)^2 + 4q(5-p)}{(p-q)(9-p)^2(7-p)} \mathfrak S_3 + \frac{4}{(9-p)(7-p)} \mathfrak S_5 + \frac{2}{(9-p)(7-p)} \mathfrak S_{\phi1}, \label{eq: Navier-Stokes 0}\\
  \frac{1}{r_H^{(p-3)/2}} \partial_i r_H^{(1)} =&\; \frac{[4(p-3)^2 + 4q(5-p)]  \mathbf v_4 + 16(p-q)  \mathbf v_5}{(p-q-1)(9-p)(7-p)(5-p)} \cr
  & + \frac{2(p-1)(p^2 - 22p + 77) - 2q(p^2 - 22p + 85)}{(p-q)(9-p)(7-p)(5-p)^2} \mathfrak V_1 \cr
  & - \frac{2(19 - 3p)}{(9-p)(7-p)(5-p)} \mathfrak V_2 - \frac{2(p^2 - 14p + 77)}{(9-p)(7-p)(5-p)^2} \mathfrak V_3 \cr
  & - \frac{2}{(7-p)(5-p)} \mathfrak V_\phi. \label{eq: Navier-Stokes i}
\end{align}
Whereas $p-q = 1$, i.e. the spatial dimension of the boundary fluid is 1, the Navier-Stokes equation becomes
\begin{align}
  \frac{1}{r_H^{(p-3)/2}} \partial_0 r_H^{(1)} =&\; \frac{16}{(9-p)^2(7-p)} \mathfrak S_3 + \frac{2}{(9-p)(7-p)} \mathfrak S_{\phi1}, \\
  \frac{1}{r_H^{(p-3)/2}} \partial_i r_H^{(1)} =&\; \frac{16}{(9-p)(7-p)(5-p)} \mathbf{v}_3 - \frac{16(p-1)}{(9-p)(7-p)(5-p)^2} \mathfrak V_1 \cr
  & - \frac{2}{(7-p)(5-p)} \mathfrak V_\phi.
\end{align}
In the above the ranges for $p$ and $q$ are $1 \leq p \leq 4$ and $q = p-1$.

\subsection{The stress-energy tensor and driving force of the fluid}

The stress-energy tensor of the dual fluid for the reduced compactified Dp-brane can be obtained by calculating the boundary limit of the following surface tensor
\begin{align}\label{eq: surface tensor cDp}
  T_{MN} =&\; \frac{1}{\kappa_{p-q+2}^2} \lim_{r\to\infty} \left( \frac rL \right)^\frac{(9-p)(p-q-1)}{2(p-q)} \bigg[ K_{MN} - h_{MN}K - \frac{9-p}{2L} \left( \frac rL \right)^{-\frac{(p-3)^2 + q(5-p)}{2(p-q)}} h_{MN} \cr
 &  - \frac{L}{4} \left( \frac rL \right)^\frac{(p-3)^2 + q(5-p)}{2(p-q)} \left( \bar{\nabla}_M \Phi \bar{\nabla}_N \Phi - \frac12 h_{MN} \left( \bar{\nabla} \Phi \right)^2 \right) \bigg].
\end{align}
With the second-order expanded on-shell metric \cref{eq: 2nd order expanded metric cDp} and the second-order solution of $\Phi$ \cref{eq: solution of Phi(2) cDp}, we get
\begin{align}\label{eq: stress-energy tensor cDp}
  T_{\mu\nu} =&\; \frac{1}{2 \kappa_{p-q+2}^2} \Bigg\{ {r_H^{7-p} \over L^{8-p}} \left( \frac{9-p}{2} u_\mu u_\nu + \frac{5-p}{2} P_{\mu\nu} \right) \cr
  & - \left( \frac{r_H}{L} \right) ^\frac{9-p}{2} \bigg( 2\sigma_{\mu\nu} + \frac{2(p-3)^2 + 2q(5-p)}{(p-q)(9-p)} P_{\mu\nu} \partial u \bigg) \cr
  & + \frac{r_H^2}{L} \Bigg[ \bigg( \frac{1}{5-p} + \frac{1}{7-p} H_\frac{5-p}{7-p} \bigg)\cdot 2\bigg( \sideset{_\langle}{}{\mathop D}\sigma_{\mu\nu\rangle} + \frac1{p-q} \sigma_{\mu\nu} \partial u \bigg) \cr
  & + \bigg( \frac{3(p-3)^2 + 3q(5-p)}{(5-p)(9-p)} - \frac{(p-3)^2 + q(5-p)}{(7-p)(9-p)} H_\frac{5-p}{7-p} \bigg) \frac{2\sigma_{\mu\nu} \partial u}{p-q} \cr
  & + \frac{1}{5-p} \cdot 4\sigma_{\langle\mu}^{~~\rho}\sigma_{\nu\rangle\rho} + \bigg( - \frac{2}{5-p} + \frac{2}{7-p} H_\frac{5-p}{7-p} \bigg) \cdot 2 \sigma_{\langle\mu}^{~~\rho} \Omega_{\nu\rangle\rho} - \frac12 \partial_{\langle\mu} \phi \partial_{\nu\rangle} \phi \Bigg] \cr
  & + P_{\mu\nu} \frac{r_H^2}{L_p} \Bigg[ \bigg( \frac{2(p-3)^2 + 2q(5-p)}{(p-q)(5-p)(9-p)} + \frac{2(p-3)^2 + 2q(5-p)}{(p-q)(7-p)(9-p)} H_\frac{5-p}{7-p} \bigg) D(\partial u) \cr
  & + \Bigg( \frac{[2(p-3)^2 + 2q(5-p)] [(3p^2 - 17p + 18) + 3q(5-p)]}{(p-q)^2 (5-p)(9-p)^2} \cr
  & + \frac{(5-p) [2(p-3)^2 + 2q(5-p)]}{(p-q) (7-p) (9-p)^2} H_\frac{5-p}{7-p} \Bigg) (\partial u)^2 \cr
  & + \frac{(p-3)^2 + q(5-p)}{(p-q)(5-p)(9-p)} \cdot 4 \sigma_{\alpha\beta}^2 - \frac{(p-3)^2+q(5-p)}{2(9-p)(p-q)} (\partial_\perp \phi)^2 \Bigg] \Bigg\}.
\end{align}
We can read the transport coefficients associated with the external scalar field as
\begin{align}
  \lambda_\phi = \frac{1}{2\kappa_{p-q+2}^2} \left( - \frac12 \right) \frac{r_H^2}{L}, \qquad  \xi_{\phi} = \frac{1}{2\kappa_{p-q+2}^2} \bigg[- \frac{(p-3)^2+q(5-p)}{2(9-p)(p-q)} \bigg] \frac{r_H^2}{L}.
\end{align}
The relation between these two is still $\xi_\phi = \frac12 \gamma^2 \lambda_\phi$, the same as the case of the reduced AdS black hole.

Following the method in \cref{subsec: deriving the force}, we have the holographic definition of the Lagrangian density for the reduced compactified Dp-brane
\begin{align}\label{eq: Lagrangian definition cDp}
  e^{-\phi} \mathcal L = \frac{1}{2 \kappa_{p-q+2}^2} \lim_{r\to\infty} \left( \frac rL \right)^\frac{(9-p)(p-q+1)}{2(p-q)} \bigg[ - \nabla_n \Phi - \frac{L}{2} \left( \frac rL \right)^\frac{(p-3)^2+q(5-p)}{2(p-q)} \bar \nabla^2 \Phi \bigg].
\end{align}
The result turns out to be
\begin{align}
  e^{-\phi} \mathcal{L} =&\; \frac{1}{2 \kappa_{p-q+2}^2}\Bigg\{ -\left( \frac{r_H}{L} \right)^\frac{9-p}{2} D \phi + \frac{r_H^2}{L} \Bigg[ \bigg( \frac{3-p}{2(5-p)} + \frac{1}{7-p}H_{\frac{5-p}{7-p}} \bigg) u^\mu u^\nu \partial_\mu \partial_\nu \phi \cr
  & - \frac12 \partial^2_{\bot} \phi + \bigg( -\frac{3}{9-p} + \frac{5-p}{(7-p)(9-p)} H_{\frac{5-p}{7-p}} \bigg) \partial u D \phi \cr
  & + \bigg( \frac{1}{5-p} + \frac{1}{7-p} H_{\frac{5-p}{7-p}} \bigg) D u^{\mu} \partial_{\mu} \phi \Bigg] \Bigg\}.
\end{align}
The coefficients now can be read as
\begin{align}
  \zeta_\phi  &= \frac1{2\kappa_{p-q+2}^2} \left( \frac{r_H}{L} \right)^\frac{9-p}{2}, \hspace{10mm} \xi_{\phi1} = \frac1{2\kappa_{p-q+2}^2} \left[ \frac{3-p}{2(5-p)} + \frac{1}{7-p}H_{\frac{5-p}{7-p}} \right] \frac{r_H^2}{L}, \cr
  \xi_{\phi2} &= \frac1{2\kappa_{p-q+2}^2} \left( - \frac12 \right) \frac{r_H^2}{L}, \qquad \xi_{\phi3} = \frac1{2\kappa_{p-q+2}^2} \left[ -\frac{3}{9-p} + \frac{5-p}{(7-p)(9-p)} H_{\frac{5-p}{7-p}} \right] \frac{r_H^2}{L}, \cr
  \xi_{\phi4} &= \frac1{2\kappa_{p-q+2}^2} \left[ \frac{1}{5-p} + \frac{1}{7-p} H_{\frac{5-p}{7-p}} \right] \frac{r_H^2}{L}.
\end{align}

Just like the reduced AdS black hole, the two identities among the second-order coefficients of the driving force are
\begin{align}
  \xi_{\phi3} = c_s^2 \xi_{\phi1} + \xi_{\phi_2}, \qquad \qquad \xi_{\phi4} = \xi_{\phi1} - \frac{p-1}{5-p} \xi_{\phi_2},
\end{align}
where $c_s^2 = \frac{5-p}{9-p}$ is the sound speed of the dual fluid for the reduced compactified Dp-brane \cite{Wu1807,Wu2012}. Note that the first identity $\xi_{\phi3} = c_s^2 \xi_{\phi1} + \xi_{\phi_2}$ takes on the same form for both the reduced AdS black hole and the reduced compactified Dp-brane. The relations between the driving-force coefficients and the stress-tensor coefficients for the case now are
\begin{align}
  \zeta_\phi = \eta, \qquad \xi_{\phi2} = \lambda_\phi, \qquad \xi_{\phi3} = - \frac{2c_s^2}{(p-q) \gamma^2} \eta \tau_\pi^*, \qquad \xi_{\phi4} = \eta \tau_\pi.
\end{align}
The above four relations all have exactly the same form as the case of reduced AdS black hole, just like the identities $\xi_\phi = \frac12 \gamma^2 \lambda_\phi$ and $\xi_{\phi3} = c_s^2 \xi_{\phi1} + \xi_{\phi_2}$.

\section{The reduced smeared Dp-brane}

Previous research has already recovered the truth that the dual fluid of the reduced smeared Dp-brane has a one-to-one correspondence to the reduced compactified Dp-brane (with $q\geq 1$) \cite{Wu2307}. Not surprisingly, all the results related to the external scalar field should also follow this rule. The full action for the reduced smeared Dp-brane is
\begin{align}\label{eq: action in CR form sDp}
  S =&\, \frac1{2\kappa_{p+2}^2} \int d^{p+2}x \sqrt{-g} \left[ R - \frac12 (\partial \varphi)^2 - V(\varphi) - \frac12 (\partial \Phi)^2 \right] \cr
  & - \frac1{\kappa_{p+2}^2} \int d^{p+1}x \sqrt{-h} \left[ K + \frac{9-p-q}{2L} \, e^{- \frac{\gamma}{2} \varphi} - \frac{L}{8} e^{\frac\gamma2 \varphi} (\bar \partial \Phi)^2 \right],
\end{align}
in which the potential of $\varphi$ and $\gamma$ are
\begin{align}
  V(\varphi) = - \frac{(7-p-q)(9-p-q)}{2L^2} e^{- \gamma \varphi}, \qquad \gamma^2 = \frac{2(p-3)^2 + 2q(p-1)}{p(9-p-q)}.
\end{align}
The EOMs derived from the above action have the same form as \cref{eq: Einstein equation,eq: EOM varphi,eq: EOM Phi}, their solutions are
\begin{align}\label{eq: background solution sDp}
  ds^2 =& \left( \frac r{L} \right)^\frac{9-p-q}{p} \left( -f(r) dt^2 + d \vv x^2 + \frac{L^{7-p-q} dr^2}{r^{7-p-q} f(r)} \right), \quad e^\varphi = \left( \frac{r}{L} \right)^{\frac{9-p-q}{2} \gamma}, \quad \Phi = \phi_0.
\end{align}
The emblackening factor in the metric above reads $f(r) = 1 - r_H^{7-p-q}/r^{7-p-q}$. From \cref{eq: surface tensor in the bulk general form}, we know the surface tensor for the reduced smeared Dp-brane is
\begin{align}\label{eq: surface tensor sDp}
  T_{MN} =&\; \frac{1}{\kappa_{p+2}^2} \lim_{r\to\infty} \left( \frac rL \right)^\frac{(9-p-q)(p-1)}{2p} \bigg[ K_{MN} - h_{MN}K - \frac{9-p-q}{2L} \left( \frac rL \right)^{-\frac{(p-3)^2 + q(p-1)}{2p}} h_{MN} \cr
 &  - \frac{L}{4} \left( \frac rL \right)^\frac{(p-3)^2 + q(p-1)}{2p} \left( \bar{\nabla}_M \Phi \bar{\nabla}_N \Phi - \frac12 h_{MN} \left( \bar{\nabla} \Phi \right)^2 \right) \bigg].
\end{align}
And by the same token as \cref{subsec: deriving the force}, one has the holographic definition of the Lagrangian for the reduced smeared Dp-brane
\begin{align}\label{eq: Lagrangian definition sDp}
  e^{-\phi} \mathcal L = \frac{1}{2 \kappa_{p+2}^2} \lim_{r\to\infty} \left( \frac rL \right)^\frac{(9-p-q)(p+1)}{2p} \bigg[ - \nabla_n \Phi - \frac{L}{2} \left( \frac rL \right)^\frac{(p-3)^2+q(p-1)}{2p} \bar \nabla^2 \Phi \bigg].
\end{align}
The action, the background, the holographic definitions of the stress tensor and the Lagrangian density are the key elements that determine the final results. From \cref{eq: action in CR form sDp,eq: background solution sDp,eq: surface tensor sDp,eq: Lagrangian definition sDp}, we can see that these elements for the reduced smeared Dp-brane can be got from those of the reduced compactified Dp-brane's by setting $p \to p+q$ in \cref{eq: action in CR form cDp,eq: background solution cDp,eq: surface tensor cDp,eq: Lagrangian definition cDp}. This coincidence suggests that the coefficients from both the stress-energy tensor and the driving force should also have such a connection between the compactified Dp-brane and the smeared Dp-brane.

Unsurprisingly, the coefficients related to the external scalar field for the reduced smeared Dp-brane turn out to be
\begin{align}
  \lambda_\phi &= \frac{1}{2\kappa_{p+2}^2} \left( - \frac12 \right) \frac{r_H^2}{L}, \qquad\;\,  \xi_{\phi} = \frac{1}{2\kappa_{p+2}^2} \bigg[- \frac{(p-3)^2+q(p-1)}{2p(9-p-q)} \bigg] \frac{r_H^2}{L}, \cr
  \zeta_\phi  &= \frac1{2\kappa_{p+2}^2} \left( \frac{r_H}{L} \right)^\frac{9-p-q}{2}, \qquad\!\! \xi_{\phi1} = \frac1{2\kappa_{p+2}^2} \left[ \frac{3-p-q}{2(5-p-q)} + \frac{1}{7-p-q}H_{\frac{5-p-q}{7-p-q}} \right] \frac{r_H^2}{L}, \cr
  \xi_{\phi2} &= \frac1{2\kappa_{p+2}^2} \left( - \frac12 \right) \frac{r_H^2}{L}, \qquad \xi_{\phi3} = \frac1{2\kappa_{p+2}^2} \left[ -\frac{3}{9-p-q} + \frac{5-p-q}{(7-p-q)(9-p-q)} H_{\frac{5-p-q}{7-p-q}} \right] \frac{r_H^2}{L}, \cr
  \xi_{\phi4} &= \frac1{2\kappa_{p+2}^2} \left[ \frac{1}{5-p-q} + \frac{1}{7-p-q} H_{\frac{5-p-q}{7-p-q}} \right] \frac{r_H^2}{L}.
\end{align}
Note some of these coefficients contain $q$, but the $q$ in the above is not the number of compactified directions but the number of smeared directions of the Dp-brane. The identities among the second-order driving-force coefficients are
\begin{align}
  \xi_{\phi3} = c_s^2 \xi_{\phi1} + \xi_{\phi_2}, \qquad \qquad \xi_{\phi4} = \xi_{\phi1} - \frac{p+q-1}{5-p-q} \xi_{\phi_2},
\end{align}
in which $c_s^2 = \frac{5-p-q}{9-p-q}$ is the sound speed for the dual fluid of the reduced smeared Dp-brane \cite{Wu2307}. These results tell the following relations
\begin{align}
  \xi_\phi = \frac12 \gamma^2 \lambda_\phi, \qquad \zeta_\phi = \eta, \qquad \xi_{\phi2} = \lambda_\phi, \qquad \xi_{\phi3} = - \frac{2c_s^2}{p \gamma^2} \eta \tau_\pi^*, \qquad \xi_{\phi4} = \eta \tau_\pi.
\end{align}
The results of $\eta$, $\eta \tau_\pi$ and $\eta \tau_\pi^*$ can be found in \cite{Wu2307}, or can also be got by setting $p \to p+q$ in \cref{eq: stress-energy tensor cDp}.

\section{Discussions and outlook}

We have explored the forced non-conformal relativistic fluid which corresponds to the Chamblin-Reall gravity with one dilaton and a manually added bulk scalar probe. We get two transport coefficients in the stress-energy tensor and five coefficients from the driving force which are related to the external scalar field. With an elaborate examination of these seven newly discovered coefficients, we also find seven identities associated with these coefficients. The seven identities can be classified into three groups:

The first group has only one identity:
\begin{align}\label{eq: identities of xi_phi and lambda_phi}
  \xi_\phi = \frac12 \gamma^2 \lambda_\phi.
\end{align}
This identity reflects the fact that the viscous term of $\xi_\phi$ can be seen as the trace of that of $\lambda_\phi$, just like the relations between the dynamical transport coefficients $\zeta = \gamma^2 \eta$ and $\xi_1 = \frac12 \gamma^2 \lambda_1$.

The second group is about the four second-order coefficients from the driving force:
\begin{align}\label{eq: identities of force coefficients}
  \xi_{\phi3} = c_s^2 \xi_{\phi1} + \xi_{\phi_2}, \qquad \qquad \xi_{\phi4} = \xi_{\phi1} - c \xi_{\phi_2},
\end{align}
in which $c$ is $p-2$ or $\frac{p-1}{5-p}$ for the reduced AdS black hole or the reduced compactified Dp-brane, respectively. The above two identities reflect the fact that only two of the second-order coefficients from the driving force are independent, which coincides with the number of coefficients in the conformal case studied in \cite{Ashok1309,Bhattacharyya0806}.

The third group contains the four relations between the coefficients of the driving force and those from the stress tensor:
\begin{align}\label{eq: identities between force and stress-tensor coefficients}
  \zeta_\phi = \eta, \qquad \xi_{\phi2} = \lambda_\phi, \qquad \xi_{\phi3} = - \frac{2c_s^2}{(p-q) \gamma^2} \eta \tau_\pi^*, \qquad \xi_{\phi4} = \eta \tau_\pi.
\end{align}
The forms of these seven identities for the reduced smeared Dp-brane can be easily obtained from the reduced compactified Dp-brane by setting $p \to p+q$. In these newly found seven identities about the seven coefficients associated with the external scalar field, only $\xi_{\phi4} = \xi_{\phi1} - c \xi_{\phi_2}$ takes on different forms for different bulk metrics. The other six have the same looking for both the reduced AdS black hole, the reduced Dp- and compactified Dp-brane, as well as the reduced smeared Dp-brane.

Through this work, we know the fluid/gravity correspondence not only be appropriate to hold for the non-conformal Chamblin-Reall background, but also be valid when the manually added bulk scalar probe exists. Meanwhile, the dual fluid is affected by an external scalar field such that a driving force appears in the Navier-Stokes equation. This inspires us to continue to study the effect of the external electromagnetic field on the non-conformal relativistic fluid dual to the Chamblin-Reall gravity. What's more, one may also add the scalar probe in the reduced D0-D4 background \cite{Wu1608}, which will lead to the coefficients associated with the viscous terms composed by derivatives of both the external scalar field and chemical potential \cite{Blake1505,Blake1507}.

At last, we find that the studies on the forced fluid are not many, except for \cite{Ashok1309,Bhattacharyya0806}, \cite{Cai1202} also explores the holographic forced non-relativistic fluid at a finite cutoff surface. We hope the information that this work uncovers can lead us to a deeper understanding of this subject.

\section*{Acknowledgement}

We would like to thank Yuye Cheng for helpful discussions at the early stage of this work. We also thank the Young Scientists Fund of the National Natural Science Foundation of China (Grant No. 11805002) for support.

\appendix

\section{The derivation of the surface tensor}

We take our boundary as $d$-dimensional, $d = p-q+1$ for the reduced compactified AdS black hole and the reduced compactified Dp-brane; while $d = p+1$ for the reduced smeared Dp-brane. The boundary parts of the full actions \cref{eq: action in CR form rAdS,eq: action in CR form cDp,eq: action in CR form sDp} can be written as
\begin{align}
  S_d &= \frac{1}{\kappa_{d+1}^2} \int d^d x \sqrt{-h} \bigg[ - K - \frac{c_1}{L} e^{- \frac\gamma2 \varphi} + c_2 L e^{\frac\gamma2 \varphi} (\bar \partial \Phi)^2 \bigg] \cr
  &\to \int d^d x \sqrt{-h} \bigg[ - 2K - \frac{2 c_1}{L} e^{- \frac\gamma2 \varphi} + 2 c_2 L e^{\frac\gamma2 \varphi} (\bar\partial \Phi)^2 \bigg].
\end{align}
Here we set $2\kappa_{d+1}^2 = 1$. The two parameters $c_{1,2}$ characterize each background and are listed in \cref{tab: c1 c2 of different models}. We have also changed the sign of the counter term $\frac{c_1}{L} e^{- \frac\gamma2 \varphi}$, comparing with the conventions in \cite{Wu1508,Wu1604,Wu1807,Wu2012,Wu2111,Wu2307}.
\begin{table}[h!]
\centering
\begin{tabular}{|c|c|c|c|}
  \hline
               & \makecell[c]{reduced AdS \\ black hole} & \makecell[c]{reduced compactified \\ Dp-brane} & \makecell[c]{reduced smeared \\ Dp-brane} \\ \hline
  \makecell[c]{dimension of \\ dual fluid $d$}   & $p-q+1$ & $p-q+1$ & $p+1$ \\ \hline
  $c_1$ & $p$ & $\frac{9-p}{2}$ & $\frac{9-p-q}{2}$ \\  \hline
  $c_2$ & $\frac{1}{4 (p-1)}$ & $\frac18$ & $\frac18$ \\
  \hline
\end{tabular}
\caption{\label{tab: c1 c2 of different models} The dimension of dual fluid and the values of $c_{1,2}$ for each model of the Chamblin-Reall gravity with one background scalar field.}
\end{table}

For the sake of simplicity, we will use Greek indices such as ``$\mu$" and ``$\nu$" for bulk indices in the following texts. Supposing the hypersurface is at $r = \text{const.}$, we then define the unit normal vector of the hypersurface as
\begin{align}
  n_\mu = \frac{\nabla_\mu r}{\sqrt{g^{\rho\sigma} \nabla_\rho r \nabla_\sigma r}},
\end{align}
whose variations are
\begin{align}
  \delta n_\mu = - \frac12 n_\mu n_\rho n_\sigma \delta g^{\rho\sigma}, \qquad \qquad \delta n^\mu = \delta g^{\mu\nu} n_\nu - \frac12 n^\mu n_\rho n_\sigma \delta g^{\rho\sigma}.
\end{align}
Taking the variation for the fields on the hypersurface, one has
\begin{align}
  \delta S_d =& \int d^d x \sqrt{-h} n^\rho g^{\mu\nu} ( \nabla_\nu \delta g_{\mu\rho} - \nabla_\rho \delta g_{\mu\nu} ) \cr
   & + \int d^d x \bigg[ \sqrt{-h} \Big( -2 K_{\mu\nu} \delta h^{\mu\nu} - 2h^{\mu\nu} \delta K_{\mu\nu} + 2 c_2 e^{\frac\gamma2 \varphi} \delta h^{\mu\nu} \bar\partial_\mu \Phi \bar\partial_\nu \Phi \Big) \cr
   & - \frac12 \sqrt{-h} h_{\mu\nu} \delta h^{\mu\nu} \bigg( -2K - \frac{2c_1}{L} e^{- \frac\gamma2 \varphi} + 2c_2 L e^{\frac\gamma2 \varphi} (\bar \partial \Phi)^2 \bigg) \bigg].
\end{align}
The first line of the above comes from the term $g^{\mu\nu} \delta R_{\mu\nu}$, whose origin is the variation of the bulk action. It can be recast like
\begin{align}
  n^\rho g^{\mu\nu} ( \nabla_\nu \delta g_{\mu\rho} - \nabla_\rho \delta g_{\mu\nu} ) &= n^\rho ( h^{\mu\nu} + n^\mu n^\nu ) ( \nabla_\nu \delta g_{\mu\rho} - \nabla_\rho \delta g_{\mu\nu} ) \cr
  &= n^\rho h^{\mu\nu} ( \nabla_\nu \delta g_{\mu\rho} - \nabla_\rho \delta g_{\mu\nu} ).
\end{align}
Thus
\begin{align}
  \delta S_d =& \int d^d x \sqrt{-h} \Big[ -2h^{\mu\nu} \delta K_{\mu\nu} + n^\rho h^{\mu\nu} ( \nabla_\nu \delta g_{\mu\rho} - \nabla_\rho \delta g_{\mu\nu} ) \Big] \cr
  & + \int d^d x \sqrt{-h} \delta h^{\mu\nu} \bigg[ -2 K_{\mu\nu} + h_{\mu\nu} K + \frac{c_1}{L} e^{- \frac\gamma2 \varphi} h_{\mu\nu} \cr
  & + 2c_2 L e^{\frac\gamma2 \varphi} \bigg( \bar\partial_\mu \Phi \bar\partial_\nu \Phi - \frac12 h_{\mu\nu} (\bar\partial \Phi)^2 \bigg) \bigg].
\end{align}
The first term in the first line contains a boundary term on the $d$-dimensional boundary, similar to the term $g^{\mu\nu} \delta R_{\mu\nu}$ in the bulk. To see this point, we need to transform this term as
\begin{align}
  &\quad\, -2 h^{\mu\nu} \delta K_{\mu\nu} = -2 h^{\mu\nu} \delta ( - h_\mu^\rho \nabla_\rho n_\nu ) \cr
  &= 2h^{\mu\nu} \Big( \delta h_\mu^\rho \nabla_\rho n_\nu + h_\mu^\rho \nabla_\rho \delta n_\nu - h_\mu^\rho \delta \Gamma_{\nu\rho}^\sigma n_\sigma \Big) \cr
  &= 2h^{\mu\nu} \bigg[ ( - \delta n_\mu n^\rho - n_\mu \delta n^\rho ) n_\rho n_\nu + \bar \nabla_\mu \delta n _\nu - h_\mu^\rho n_\sigma \frac12 g^{\sigma \lambda} ( \nabla_\nu \delta g_{\rho \lambda} + \nabla_\rho \delta g_{\nu\lambda} - \nabla_\lambda \delta g_{\nu\rho} ) \bigg] \cr
  &= 2 \bar\nabla^\mu \delta n_\mu - h^{\mu\nu} n^\rho ( \nabla_\mu \delta g_{\nu\rho} + \nabla_\nu \delta g_{\mu\rho} - \nabla_\rho \delta g_{\mu\nu} ).
\end{align}
Then
\begin{align}
  &\quad\, -2 h^{\mu\nu} \delta K_{\mu\nu} + n^\rho h^{\mu\nu} ( \nabla_\nu \delta g_{\mu\rho} - \nabla_\rho \delta g_{\mu\nu} ) = 2 \bar\nabla^\mu \delta n_\mu - n^\nu \bar\nabla^\mu \delta g_{\mu\nu} \cr
  &= 2 \bar\nabla^\mu \delta n_\mu - \bar\nabla^\mu ( n^\nu \delta g_{\mu\nu} ) + ( \bar\nabla^\mu n^\nu ) \delta g_{\mu\nu} \cr
  &= \bar\nabla^\mu ( 2 \delta n_\mu - \delta n_\mu + g_{\mu\nu} \delta n^\nu ) + K_{\mu\nu} ( \delta h^{\mu\nu} + \delta n^\mu n^\nu + n^\mu \delta n^\nu ) \cr
  &= \bar\nabla^\mu ( \delta n_\mu + g_{\mu\nu} \delta n^\nu ) + K_{\mu\nu} \delta h^{\mu\nu}
\end{align}
Considering that
\begin{align}
  &~~~~ \delta n_\mu + g_{\mu\nu} \delta n^\nu \cr
  &= \delta n_\mu + n_\mu n_\nu \delta n^\nu + h_{\mu\nu} \delta n^\nu \cr
  &=  - \frac12 n_\mu n_\rho n_\sigma \delta g^{\rho\sigma} + n_\mu n_\nu \Big( \delta g^{\nu\rho} n_\rho - \frac12 n^\nu n_\rho n_\sigma \delta g^{\rho\sigma} \Big) + h_{\mu\nu} \delta n^\nu = h_{\mu\nu} \delta n^\nu,
\end{align}
thus we have
\begin{align}
  -2 h^{\mu\nu} \delta K_{\mu\nu} + n^\rho h^{\mu\nu} ( \nabla_\nu \delta g_{\mu\rho} - \nabla_\rho \delta g_{\mu\nu} ) = \bar\nabla_\mu \delta n^\mu + K_{\mu\nu} \delta h^{\mu\nu}.
\end{align}
Here we use the fact that $h_{\mu\nu}$ is compatible with the covariant derivative on the hypersurface $\bar\nabla^\mu$. In deriving this result, we can see that both the two terms in the right-hand-side of the above equation come from $-2h^{\mu\nu} \delta K_{\mu\nu}$. So $\delta S_d$ can be finally formulated as
\begin{align}
  \delta S_d =& \int d^d x \sqrt{-h} \Big( \bar\nabla_\mu \delta n^\mu + K_{\mu\nu} \delta h^{\mu\nu} \Big) + \int d^d x \sqrt{-h} \delta h^{\mu\nu} \bigg[ -2 K_{\mu\nu} + h_{\mu\nu} K \cr
  & + \frac{c_1}{L} e^{- \frac\gamma2 \varphi} h_{\mu\nu} + 2c_2 L e^{\frac\gamma2 \varphi} \bigg( \bar\partial_\mu \Phi \bar\partial_\nu \Phi - \frac12 h_{\mu\nu} (\bar\partial \Phi)^2 \bigg) \bigg] \cr
  =& \int d^d x \sqrt{-h} \bar\nabla_\mu \delta n^\mu + \int d^d x \sqrt{-h} \delta h^{\mu\nu} \bigg[ - K_{\mu\nu} + h_{\mu\nu} K + \frac{c_1}{L} e^{- \frac\gamma2 \varphi} h_{\mu\nu} \cr
  & + 2c_2 L e^{\frac\gamma2 \varphi} \bigg( \bar\partial_\mu \Phi \bar\partial_\nu \Phi - \frac12 h_{\mu\nu} (\bar\partial \Phi)^2 \bigg) \bigg].
\end{align}
The first term in the third line of the above is a boundary term on the $d$-dimensional boundary and thus can be omitted if we suppose the boundary is an infinite hypersurface. Defining the surface tensor by
\begin{align}
  T_{\mu\nu}^{S} \equiv \frac{-2}{\sqrt{-h}} \frac{\delta S_d}{\delta h^{\mu\nu}},
\end{align}
we can finally get the surface tensor in the bulk as
\begin{align}\label{eq: surface tensor in the bulk general form}
  T_{\mu\nu}^{S} &= 2 \bigg[ K_{\mu\nu} - h_{\mu\nu} K - \frac{c_1}{L} e^{- \frac\gamma2 \varphi} h_{\mu\nu} - 2c_2 L e^{\frac\gamma2 \varphi} \bigg( \bar\partial_\mu \Phi \bar\partial_\nu \Phi - \frac12 h_{\mu\nu} (\bar\partial \Phi)^2 \bigg) \bigg] \cr
  & \to \frac{1}{\kappa_{d+1}^2} \bigg[ K_{\mu\nu} - h_{\mu\nu} K - \frac{c_1}{L} e^{- \frac\gamma2 \varphi} h_{\mu\nu} - 2c_2 L e^{\frac\gamma2 \varphi} \bigg( \bar\partial_\mu \Phi \bar\partial_\nu \Phi - \frac12 h_{\mu\nu} (\bar\partial \Phi)^2 \bigg) \bigg]. \qquad
\end{align}
The bulk surface gravity $2\kappa_{d+1}^2$ has been restored in the end.


\providecommand{\href}[2]{#2}\begingroup\raggedright\endgroup

\end{document}